\documentclass[epj]{webofc}
\usepackage[utf8]{inputenc}
\usepackage[varg]{txfonts}   
\usepackage{booktabs}
\usepackage{xcolor}
\definecolor{darkred}{rgb}{0.4,0.0,0.0}
\definecolor{darkgreen}{rgb}{0.0,0.4,0.0}
\definecolor{darkblue}{rgb}{0.0,0.0,0.4}
\usepackage[bookmarks,linktocpage,colorlinks,
    linkcolor = darkred,
    urlcolor  = darkblue,
    citecolor = darkgreen]{hyperref}
\usepackage{subfigure}
\usepackage{grffile}
\usepackage{amsmath}
\wocname{EPJ Web of Conferences}
\woctitle{Lattice2017}
\newcommand{\Sp}[1]{\ensuremath{\mathrm{Sp}(#1)}}
\newcommand{\SU}[1]{\ensuremath{\mathrm{SU}(#1)}}
\newcommand{\Nf}{N_{\textnormal{f}}}

\newcommand\eee{\mathcal{E}}
\newcommand\www{\mathcal{W}}
\graphicspath{{figs/}}
\DeclareMathOperator{\re}{Re}
\DeclareMathOperator{\tr}{tr}
\begin{document}
%
\selectlanguage{english}
\title{%
Higgs compositeness in $\Sp{2N}$ gauge theories --- \\Resymplecticisation, scale setting and topology

}
\author{%
\firstname{Ed} \lastname{Bennett}\inst{1}\fnsep\thanks{Speaker, \email{e.j.bennett@swansea.ac.uk}}\fnsep\thanks{Funded by the Supercomputing Wales project, which is part-funded by the European Regional Development Fund (ERDF) via Welsh Government.} \and
\firstname{Deog Ki} \lastname{Hong}\inst{2}\fnsep\thanks{Supported in part by Korea Research Fellowship program funded by the Ministry of Science, ICT and Future Planning through the National Research Foundation of Korea (2016H1D3A1909283) and under the framework of international cooperation program (NRF-2016K2A9A1A01952069).} \and
\firstname{Jong-Wan}  \lastname{Lee}\inst{2}\fnsep\inst{3}\fnsep\footnotemark[3] \and
\firstname{C.-J.~David} \lastname{Lin}\inst{4}\fnsep\thanks{Supported by Taiwanese MoST grant 105-2628-M-009-003-MY4} \and \\
\firstname{Biagio} \lastname{Lucini}\inst{5}\fnsep\thanks{Supported in part by the Royal Society and the Wolfson Foundation.}\fnsep\thanks{Supported in part by the STFC Consolidated Grants ST/L000369/1 and ST/P00055X/1.} \and
\firstname{Maurizio} \lastname{Piai}\inst{6}\fnsep\footnotemark[6] \and
\firstname{Davide} \lastname{Vadacchino}\inst{1}\fnsep\footnotemark[2]\fnsep\footnotemark[6]
}
\institute{%
Swansea Academy of Advanced Computing, Swansea University, Singleton Park, Swansea, SA2 8PP, UK
\and
Department of Physics, Pusan National University, Busan 46241, Korea
\and
Extreme Physics Institute, Pusan National University, Busan 46241, Korea
\and
Institute of Physics, National Chiao-Tung University, Hsinchu 30010, Taiwan
\and
Department of Mathematics, Swansea University, Singleton Park, Swansea, SA2 8PP, UK
\and
Department of Physics, Swansea University, Singleton Park, Swansea, SA2 8PP, UK
}
\abstract{%
As part of an ongoing programme to study \Sp{2N} gauge theories as potential realisations of composite Higgs models, we consider the case of \Sp{4} on the lattice, both as a pure gauge theory, and with two Dirac fermion flavors in the fundamental representation. In order to compare results between these two cases and maintain control of lattice artefacts, we make use of the gradient flow to set the scale of the simulations. We present some technical aspects of the simulations, including preliminary results for the scale setting in the two cases and results for the topological charge history.}

\maketitle
\section{Introduction}\label{intro}
Due to its significant potential phenomenological interest, we have embarked on an extended programme to study \Sp{2N} gauge theories, as described in more detail in \cite{biagiojongwan}. In this contribution we report on some of the technical details of the study to date.

In the current, exploratory phase of the study, we have first used the heat bath algorithm to study the quenched \Sp{4} theory, including its glueball spectrum (described in \cite{davide}) and the quenched meson spectrum, which we have fitted to an effective field theory (described in \cite{biagiojongwan}). We have then implemented the Hybrid Monte Carlo (HMC) algorithm for dynamical fermions in the \Sp{4} gauge theory, which will be described in the first half of this contribution, along with a discussion of the resymplecticisation procedure used for the heat bath algorithm. An exploration of the phase structure of the dynamical theory, and preliminary results for its spectroscopy,  are presented in \cite{biagiojongwan}. We then set the scale of the theories using the gradient flow scales $t_0$ and $w_0$, and verify the ergodicity of the topological charge and observe the topological susceptibility, the discussion of which forms the second half of this contribution.

\section{Lattice action and the HMC algorithm}
In \cite{davide, biagiojongwan}, we have presented results from the pure gauge \Sp{4} theory, which we use in \cite{biagiojongwan} to perform fits to Effective Field Theory (EFT). For this purpose, we were able to use the heat bath algorithm, due to its high ergodicity and computational efficiency. However, we ultimately intend to perform simulations with dynamical fermions, in order to fully understand the behaviour near the chiral limit. In this case the heat bath algorithm is no longer sufficient, and we instead make use of the HMC algorithm. We must therefore adapt the algorithm to work with the \Sp{4} gauge group.

As in \cite{davide}, we use the Wilson gauge action; to this we add the Wilson fermion action. The molecular dynamics portion of the HMC update evolves the motion under the gauge force
\begin{equation}
	F_\textnormal{G}^A(x, \mu) = \frac{\beta}{4}\frac{1}{T_F} \re \tr_c \left[iT^A U_\mu(x) V_\mu^\dagger(x)\right]\;,
\end{equation}
where $V_\mu(x)$ is the sum of the forward and backward staples around the link $U_\mu(x)$, and $T^A$ are the generators of the \Sp{4} gauge group. The update then proceeds as detailed in \cite{DelDebbio:2008tv}.

\section{Resymplecticisation}
\begin{figure}[htb]
\centering
\subfigure[With reunitarisation only..]{\includegraphics[width=0.48\textwidth]{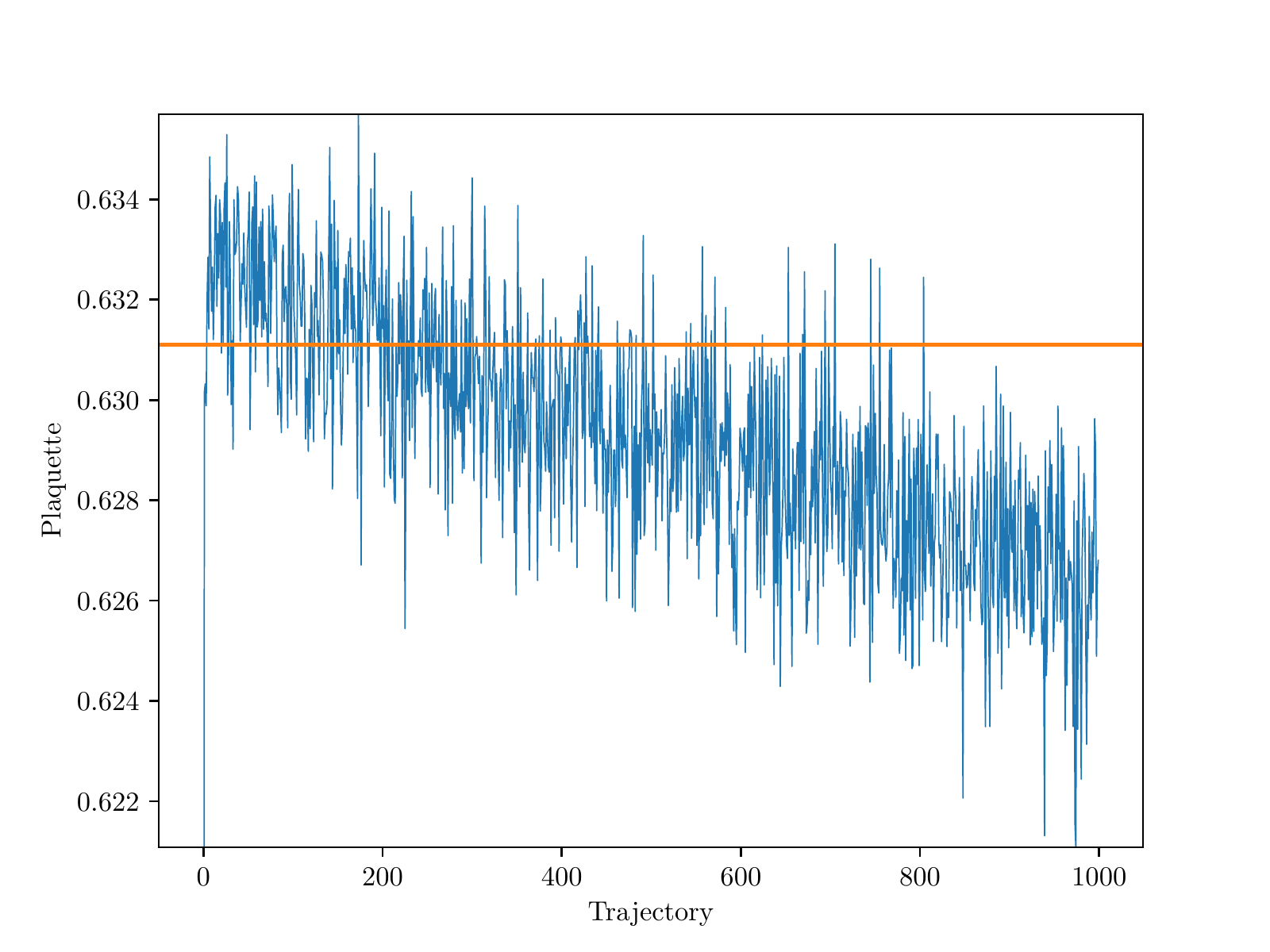}}
\subfigure[With resymplecticisation.]{\includegraphics[width=0.48\textwidth]{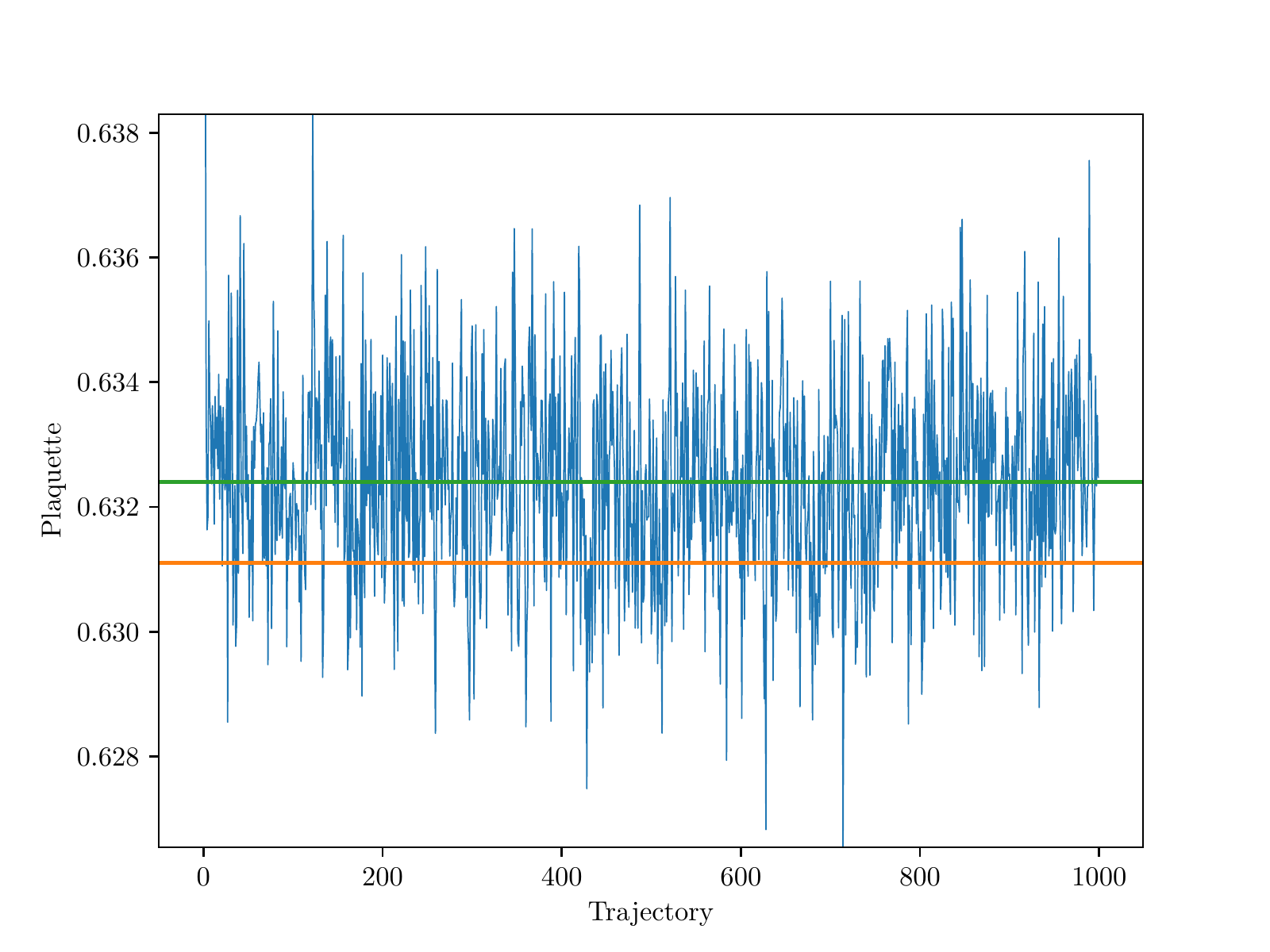}}
\caption{The time history of the plaquette for $\Nf=2$, $\beta=8.0$, $m=1.0$, with and without making use of the resymplecticisation procedure described in the text. In orange is the value obtained in \cite{Holland:2003kg}. The green line in the right panel corresponds to the average of the data set.}
\label{fig:resymp}
\end{figure}

Since the HMC algorithm is implemented in a computer which operates at finite precision, it is possible that as the simulation progresses, the group elements will diverge out of the group unless a constraint is applied. In simulations of \SU{N} theories, this constraint is known as reunitarisation. When reunitarisation is applied to the \Sp{4} theory, the result shown in Fig.~\ref{sf:noresymp} is obtained; equilibrium is reached after a handful of updates, but the simulation slowly drifts away. It is likely that if left for longer, the result would converge on that of \SU{4}. Thus we require a procedure to return the group element back to the \Sp{4} group after each update.

To do this, we represent an \Sp{4} matrix as:
\begin{align}
	Q(x, \mu) &= Q_0 (x, \mu) \otimes \mathbb{I}_2 + Q_1(x, \mu) \otimes e_1 + Q_2(x, \mu) \otimes e_2 + Q_3(x, \mu) \otimes e_3\;,
\end{align}
where
\begin{align}
	\mathbb{I}_2 &= \left( \begin{array}{cc}1 & 0 \\ 0 & 1\end{array}\right), & e_1 &= \left( \begin{array}{cc}i & 0 \\ 0 & -i\end{array}\right), & e_2 &= \left( \begin{array}{cc}0 & 1 \\ -1 & 0\end{array}\right), & e_3 &= \left( \begin{array}{cc}0 & i \\ i & 0\end{array}\right).
\end{align}
Projecting a matrix that may have deviated from \Sp{4} onto this basis will give a matrix which is once again an element of \Sp{4}. When this is done after every update, the MC history of the plaquette looks as in Fig.~\ref{sf:resymp}. In both cases the result is for heavy, but not quenched, fermions; as might be expected, the average plaquette in the stable case is close but not identical to the quenched result of \cite{Holland:2003kg}.

For the pure gauge calculations, an alternative procedure was used, derived from the (modified) Gram--Schmidt algorithm. Noting that a general \Sp{2N} matrix $U_{ij}$ is overconstrained, then if elements $U_{ij}$, $i\in[1,N]$, $\forall j$ are known, then the remainder may be calculated as
\begin{equation}
	U_{i+N,j} = -\Omega U_{ij}^*\;,
\end{equation}
where $\Omega$ is the symplectic matrix
\begin{equation}
	\Omega = \left( \begin{array}{cccc} 0 & 0 & 1 & 0 \\ 0 & 0 & 0 & 1 \\ -1 & 0 & 0 & 0 \\ 0 & -1 & 0 & 0 \end{array}\right)\;.
\end{equation}
Thus after normalising the elements $U_{1,j}$, the elements $U_{N+1,j}$ can be calculated. By orthonormalising with respect to both of these sets of elements, then $U_{2,j}$ may be calculated, and the process repeats until a full matrix is obtained, which is then guaranteed to be in $\Sp{2N}$.

\section{Gradient flow}
In order to draw comparisons between data at different values of $\beta$ and $\Nf$, it is necessary to set a scale, in order to remove the dependence on the unknown lattice spacing $a$ from results. This can be done by rescaling to common units, most easily by making use of the gradient flow, as popularised by Martin L\"uscher \cite{Luscher:2010iy}. A flow is defined starting on the gauge field $\left.B_\mu\right|_{t=0}=A_\mu$ as generated by the Monte Carlo method, and evolving under the diffusion equation $\dot{B}_\mu=D_\nu G_{\nu\mu}$, where $G_{\mu\nu}=\partial_\mu B_\nu - \partial_\nu B_\mu \left[B_\mu, B_\nu\right]$, and $D_\mu = \partial_\mu + \left[ B_\mu, \frac{\partial}{\partial t}\right]$.

\vspace{4pt}In terms of lattice observables, this takes the form 
\begin{align}
	\frac{\partial V_\mu(t,x)}{\partial t} &= -g_0^2 \left\{ \partial_{x,\mu}S_{\textnormal{latt.}}[V_\mu]\right\}V_\mu(t,x)\;&V_\mu(0,x)&=U_\mu(x)\;,
\end{align}
where $S_{\textnormal{latt.}}$ is the lattice action. This is discretised by using a fourth-order Runge--Kutta integrator, and has the effect that at $t\ne0$, observables are renormalised.

In principle, it would be possible for this discrete flow to introduce finite precision errors causing the $V_\mu(t,x)$ to leave the group; for this reason we added the resymplecticisation described in the previous section after each flow step. This was however found to have no effect on the results, so was disabled to improve the efficiency of the program.

\begin{figure}[htb]
\centering
\subfigure[$\eee(t)$\label{sf:noresymp}]{\includegraphics[width=0.48\textwidth]{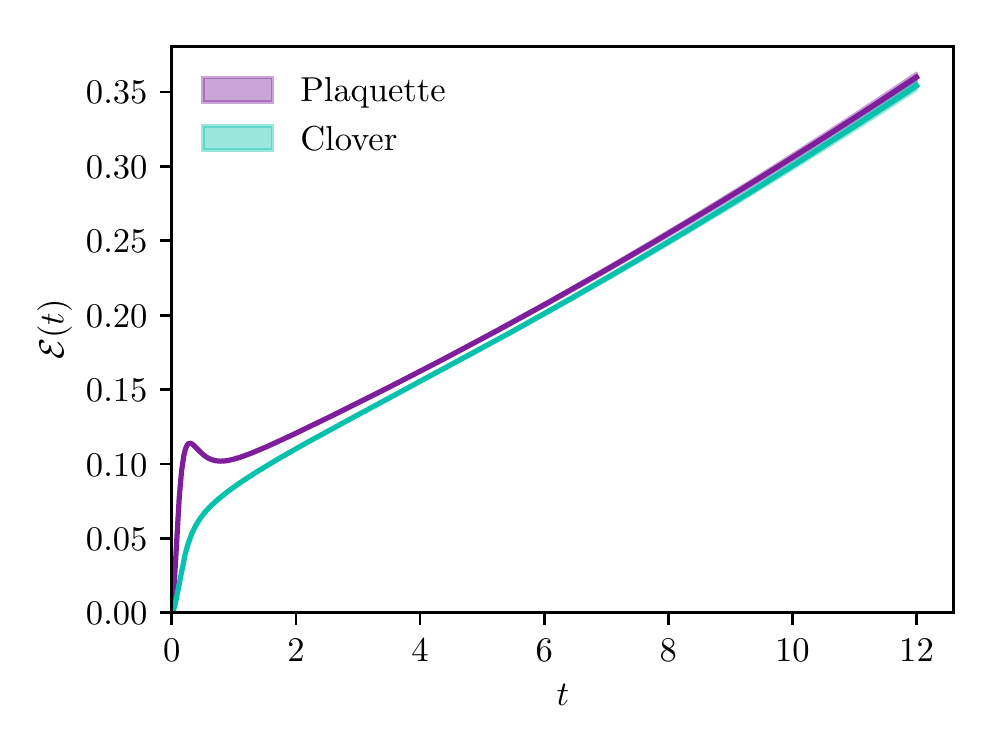}}
\subfigure[$\www(t)$\label{sf:resymp}]{\includegraphics[width=0.48\textwidth]{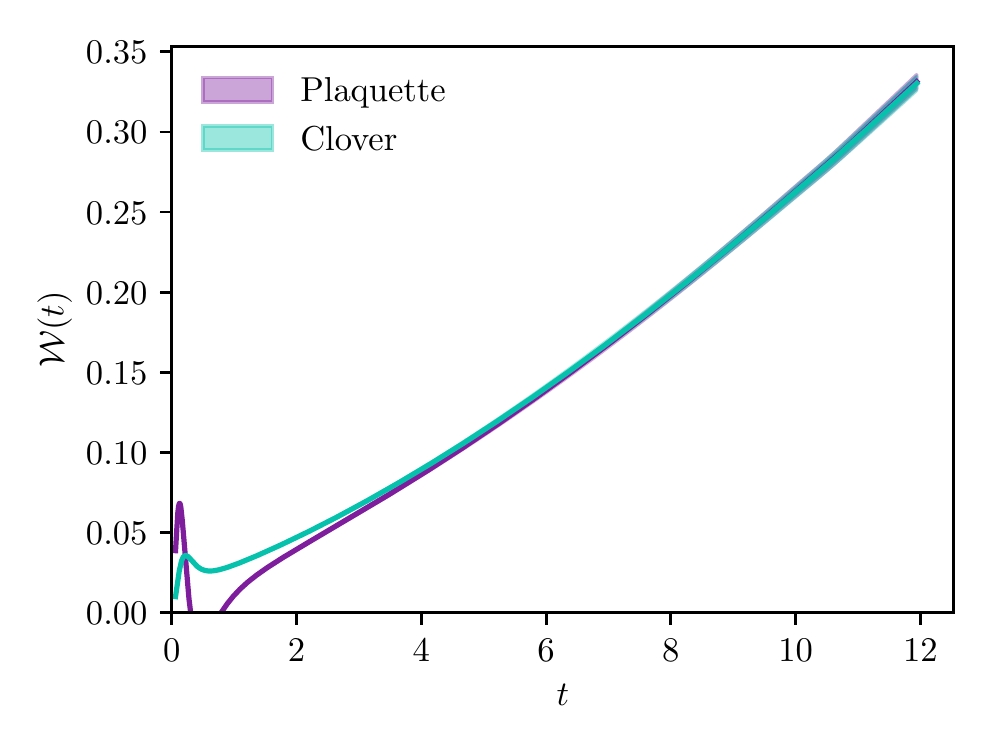}}
\caption{History of the quantities $\eee(t)$ and $\www(t)$ as a function of the flow time $t$ for the pure gauge ensemble at $\beta=8.4$, comparing the two different definitions of $\eee$ described in the text.}
\label{fig:flowhist}
\end{figure}

By considering the energy density $E=-\frac{1}{2}\tr(G_{\mu\nu}G_{\mu\nu})$, a scale $t_0$ can be extracted by finding the value of the flow time at which $\left.\langle t^2 E\rangle\right|_{t=t_0} = \eee_0$, where $\eee_0$ is a reference value to be chosen. Alternatively, as proposed in \cite{Borsanyi:2012zs}, one can take the time derivative, defining $w_0$ as $\left.t \frac{\partial}{\partial t} \langle t^2 E\rangle\right|_{t=w_0^2} = \www_0$, with $\www_0$ again a reference value to be chosen. In each case the value of $t_0$, $w_0$ has been assumed to have a very weak lattice spacing dependence in physical units, thus any change in $t_0^2 / a^2$, $w_0/a$ as a function of $a$ results from the explicit $a$-dependence in the denominator. 

\begin{figure}[htb]
\centering
\includegraphics[width=0.7\textwidth]{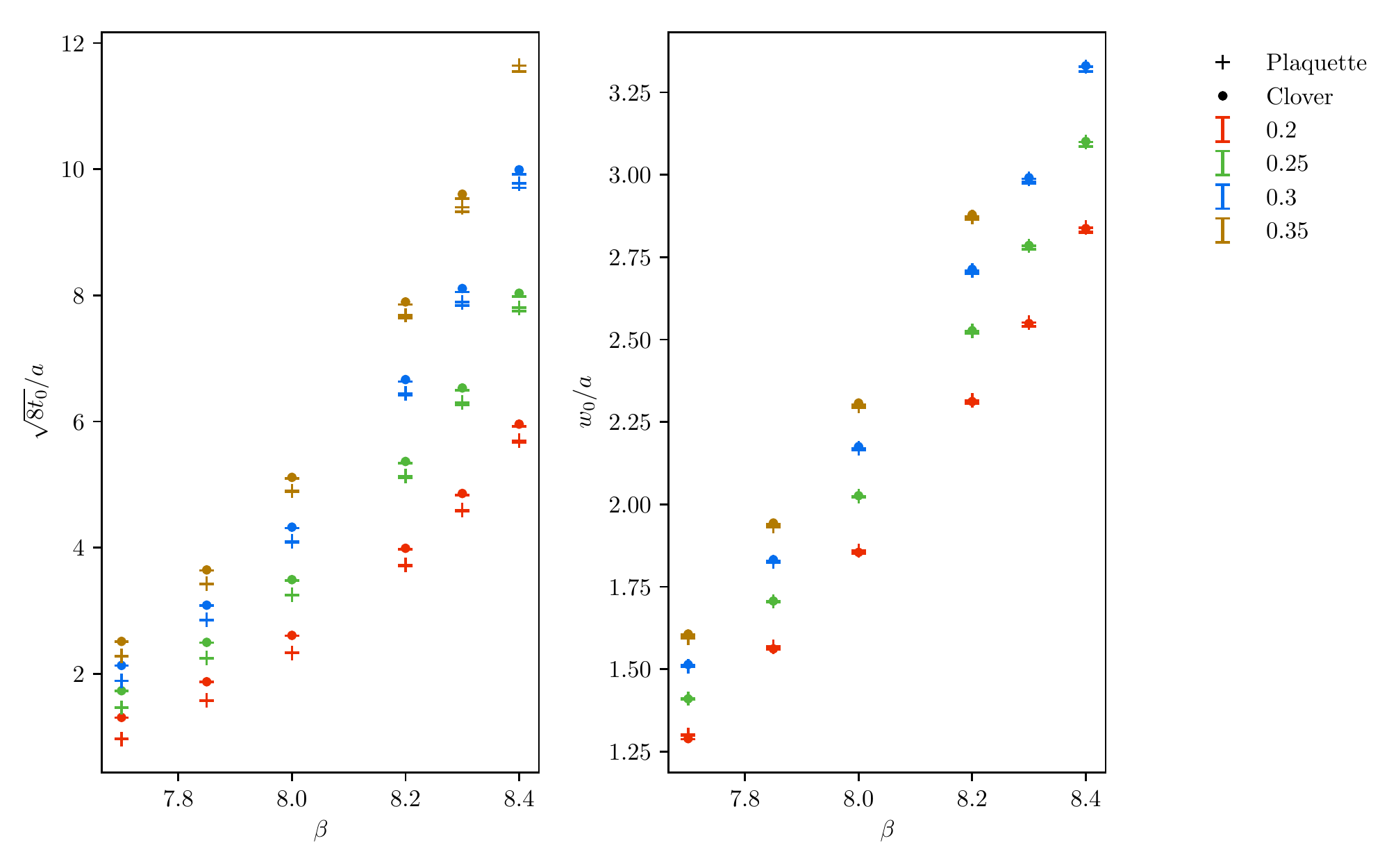}
\caption{The gradient flow scales $t_0$ and $w_0$ for the pure gauge theory, showing a variety of possible values of $\eee_0=\www_0$ as indicated in the legend.}
\label{fig:flowscalespg}
\end{figure}

\begin{figure}[htb]
\centering
\subfigure[\label{sf:2f}]{\includegraphics[width=0.7\textwidth]{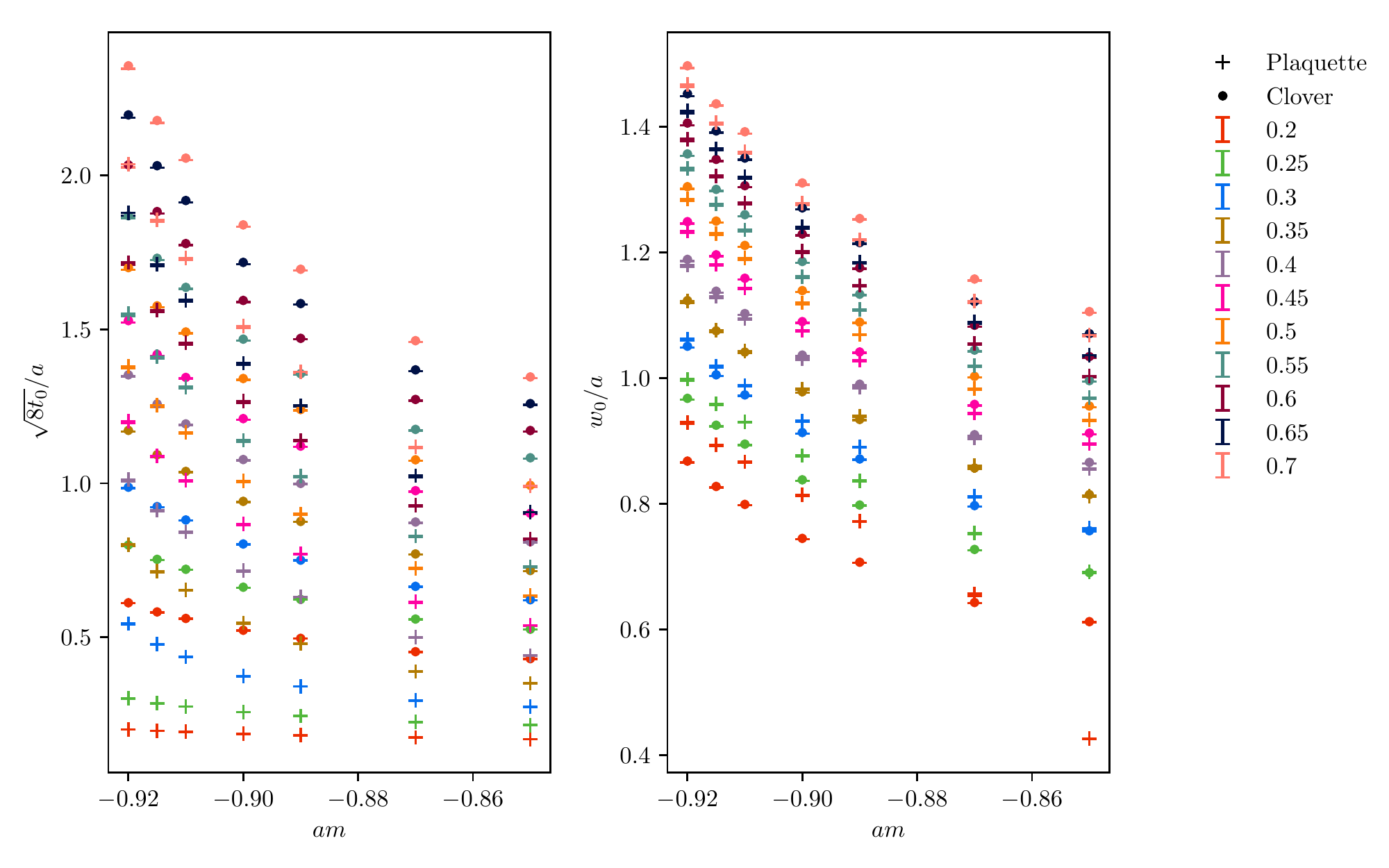}}
\subfigure[\label{sf:comp}]{\includegraphics[width=0.7\textwidth]{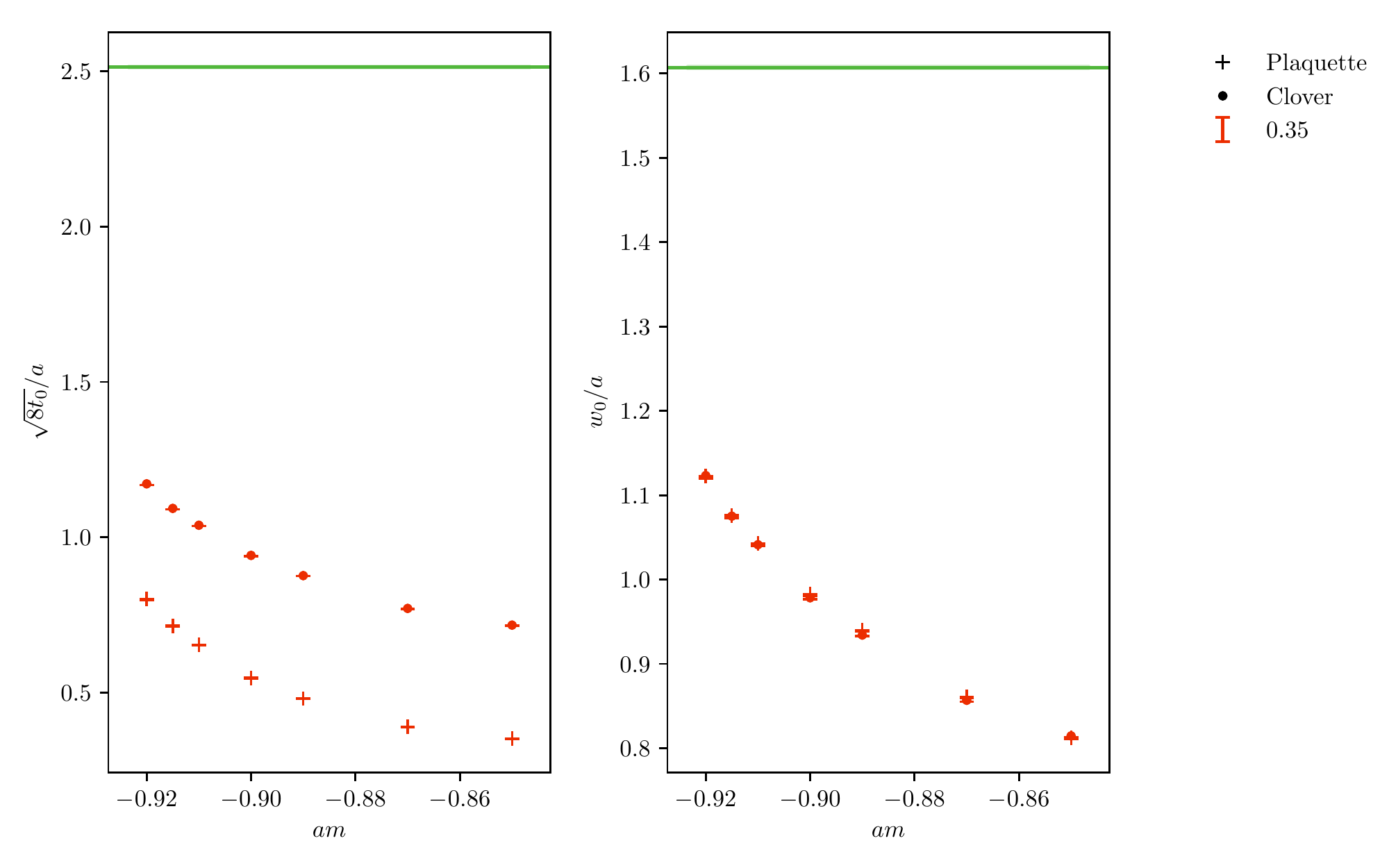}}
\caption{The gradient flow scales $t_0$ and $w_0$ for the $\Nf=2$ theory: (a) showing a variety of possible values of $\eee_0=\www_0$ as indicated in the legend, and (b) showing the single value $\eee_0=\www_0=0.35$, comparing with the same case for the pure gauge theory at $\beta=7.7$ (green line).}
\label{fig:flowscales2f}
\end{figure}

There are two commonly-used discretisations for $E$: $E_{\textnormal{plaq.}}$ via the average plaquette, and $E_{\textnormal{sym.}}$ via the symmetric four-plaquette clover. They must agree in the continuum limit, and thus discrepancies between them are indicative of finite-lattice spacing artefacts. As can be see in Fig.~\ref{fig:flowhist}, the two show an initial discrepancy but reach agreement after a short time; this has implications on our choice of $\eee_0$ and $\www_0$. This choice is explored in Figs.~\ref{fig:flowscalespg} and \ref{sf:2f}, showing how the choice affects $t_0$, $w_0$ in the pure gauge theory and the $\Nf=2$ theory respectively. In the latter, we show the closest agreement at $\eee_0=\www_0=0.35$, which we adopt for the subsequent discussion. 

While $w_0$ agrees between the two definitions of $E$ at this value, the values for $t_0$ do not. Figure~\ref{sf:comp} pulls out the detail for this choice, showing the $\Nf=2$ results at $\beta=6.9$ (red points) compared with the pure gauge result for $\beta=7.7$, the coarsest lattice considered to give reliable results \cite{davide}. Since $\sqrt{8t_0}/a$, $w_0/a$ each have a $1/a$ dependence in lattice units, lower values indicate coarser lattice spacings; thus the $\Nf=2$ results are substantially coarser than the coarsest acceptable pure gauge ensemble, explaining the divergence in the values of $t_0$. While the dynamical results presented here are proof-of-concept, for future studies of dynamical simulations, a higher value of $\beta$ will be needed to give an acceptably fine lattice.

\section{Topology}
As the lattice extent is finite in all directions, a given configuration will fall into one of a number of topological sectors, labelled by an integer (or, at finite $a$, near-integer) topological charge $Q$, which is expected to have a Gaussian distribution about zero. Since it is probabilistically unfavourable to change a discrete global observable using a small local update, $Q$ can show very long autocorrelations; as the continuum limit (i.e.\ the limit of integer $Q$) is approached, $Q$ can ``freeze'', ceasing to change at all.

It is necessary to check that $Q$ is not frozen, and instead moves ergodically, for two reasons: firstly, the exponential autocorrelation time of the Monte Carlo simulation as a whole scales as the longest autocorrelation time in the system; this is likely to be that for $Q$. Secondly, the values of physical observables depend on which topological sector a configuration is in \cite{Brower:2003yx,Galletly:2006hq}; sampling a single $Q$ or an unrepresentative distribution of $Q$s will introduce an uncontrolled systematic error. It is therefore necessary to verify that $Q$ not only moves sufficiently rapidly, but also displays the expected Gaussian histogram.

\begin{figure}[htb]
\centering
\subfigure[Pure gauge, $L=24$, $\beta=7.7$. $Q_0=-0.09\pm0.32$; $\sigma=6.32\pm0.32$.]{\includegraphics[width=0.65\textwidth]{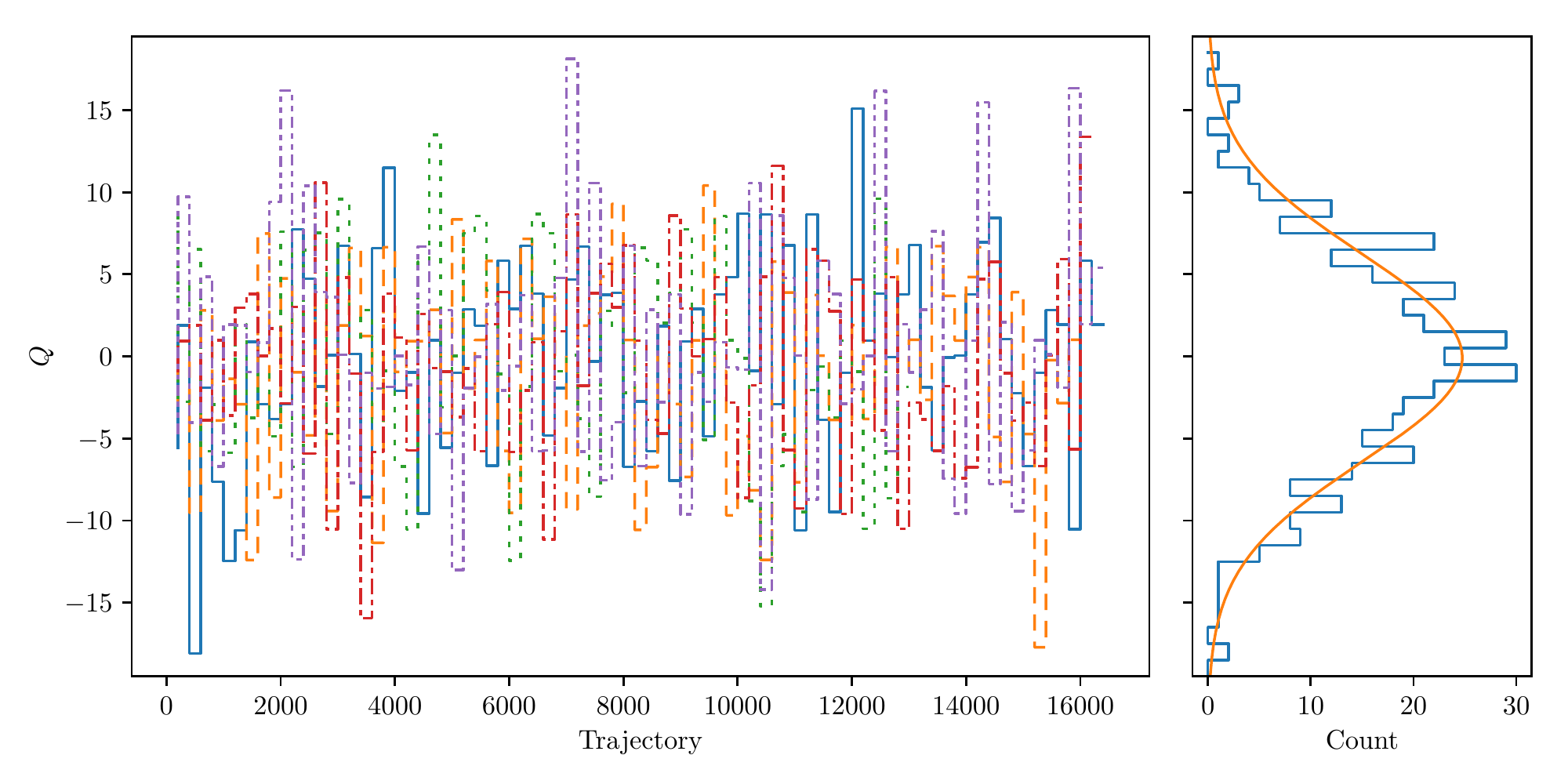}}
\subfigure[$\Nf=2$, $L=16$, $\beta=6.9$, $m=-0.92$. $Q_0=1.26\pm0.97$; $\sigma = 6.61\pm0.98$.]{\includegraphics[width=0.65\textwidth]{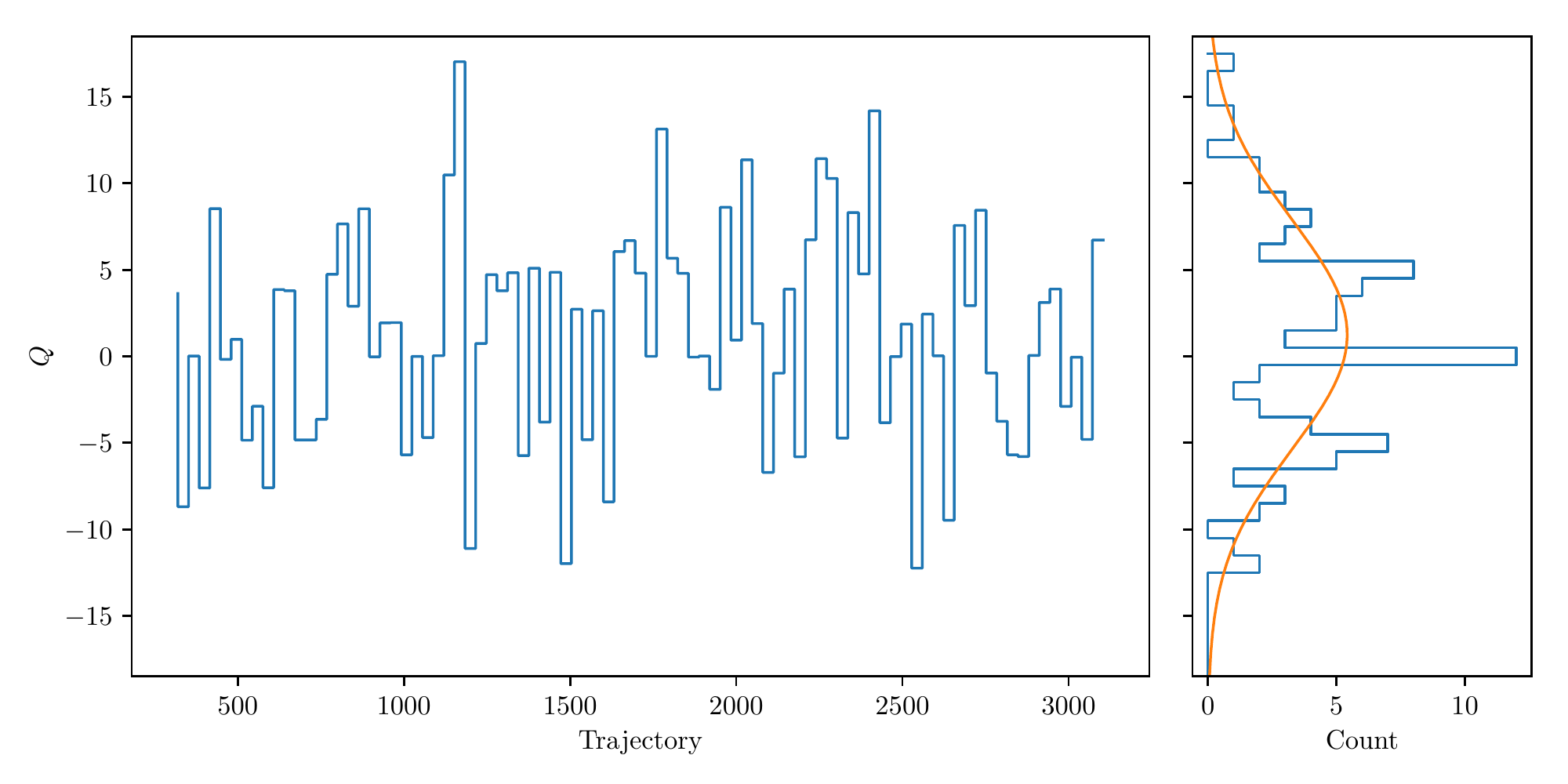}}
\caption{Topological charge histories and histograms for the specified ensembles.}
\label{fig:qhistory}
\end{figure}

$Q$ is computed on the lattice as
\begin{align}
	Q &= \sum_x q(x)\;, &&\textnormal{where} & q(x) = \frac{1}{32\pi^2} \epsilon_{\mu\nu\rho\sigma} \operatorname{tr} \left\{U_{\mu\nu}(x)U_{\rho\sigma}(x)\right\}\,,
\end{align}
and $x$ runs over all lattice sites. For gauge configurations generated by Monte Carlo studies, this observable is dominated by ultraviolet fluctuations; therefore it is necessary to perform some sort of smoothing to extract the true value. The gradient flow, as described in the previous section, is used for this purpose; the calculation of $Q$ is performed for near-zero cost as the values of $E$ are calculated.

The topological charge history was plotted for all ensembles, including both pure gauge and those with matter. In all cases, $Q$ was found to move with no noticeable autocorrelation, and showed the expected Gaussian distribution centred on $Q = 0$, modulo relatively small statistics on some dynamical ensembles. A sample of these histories are shown in Fig.~\ref{fig:qhistory}.

\begin{figure}[h!]
\centering
\includegraphics[width=0.65\textwidth]{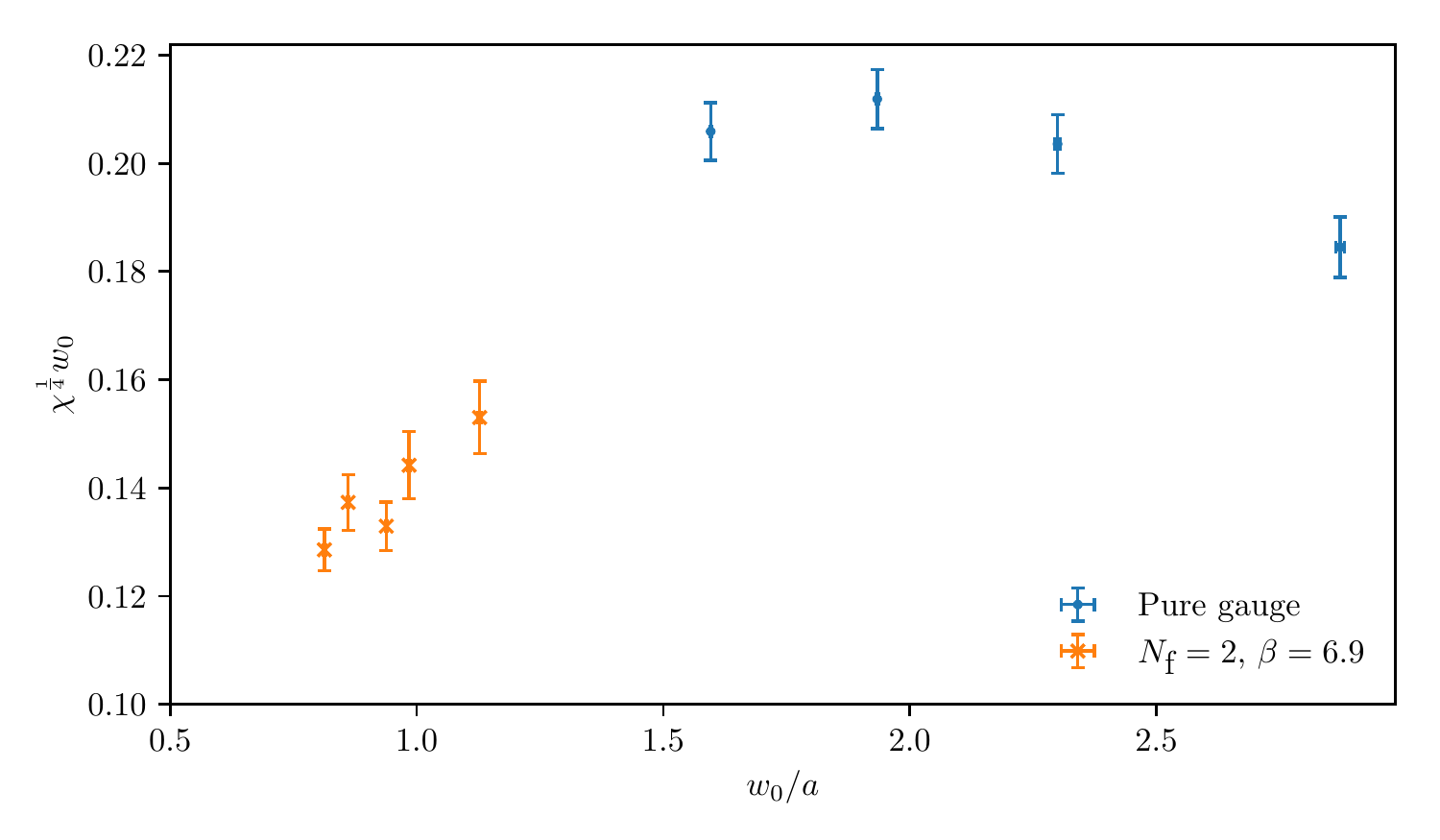}
\caption{The topological susceptibility of the pure gauge and flavoured theories as a function of the gradient flow scale $w_0$.}
\label{fig:suscept}
\end{figure}

It is also of interest to observe the topological susceptibility, defined as $\chi = (\langle Q^2\rangle - \langle Q \rangle^2) / V$; this is plotted as a dimensionless ratio with the gradient flow $w_0$, as a function of $w_0$, in Fig.~\ref{fig:suscept}. By comparison with Fig.~\ref{fig:flowscales2f}, it is visible that in the heavy fermion limit, the susceptibility does not approach that of the pure gauge theory. We interpret this as a second symptom that $\beta=6.9$ is too coarse to obtain reasonable data with $\Nf=2$ in this theory.

\section{Conclusions}
We have implemented the HMC algorithm for dynamical fermions in the \Sp{4} gauge theory, making use of a quaternion-based resymplecticisation scheme. We have also implemented a modified Gram--Schmidt-based resymplecticisation scheme for general \Sp{2N}, which we have used for heat bath computations. We have then made use of the gradient flow to study the scales $t_0$ and $w_0$, as well as the topological charge distribution and susceptibility, for both the pure gauge theory at a variety of values for $\beta$, and the dynamical theory with $\Nf=2$ at $\beta=6.9$ as an initial proof of concept. The topological charge is found to move well in all cases, although the statistics are relatively small in the $\Nf=2$ cases. The flow scale $t_0$ is found to be self-inconsistent depending on how it is calculated for $\Nf=2$, indicating that $\beta=6.9$ is too far from the continuum, a conclusion that is supported by the behaviour of the topological susceptibility. Tests of the spectroscopy of the dynamical theory are shown in \cite{biagiojongwan}; the next step in this programme of work is to expand the scope of the dynamical simulations to production runs with larger volumes and statistics, at suitably fine lattice spacings.

\section*{Acknowledgements}
We acknowledge the support of the Supercomputing Wales project, which is part-funded by the European Regional Development Fund (ERDF) via Welsh Government.

\bibliography{lattice2017}

\end{document}